\documentclass[11pt,letterpaper,notoc,nohyper]{JHEP3}
\usepackage{graphics}
\usepackage{amsmath}
\usepackage{amssymb}
\usepackage{cite}
\usepackage{colordvi}
\usepackage{feynmp}
\usepackage{epsfig}
\usepackage{latexsym}

\newcommand{\twi}{\widetilde}

\newcommand{\beq}{\begin{eqnarray}}
\newcommand{\eeq}{\end{eqnarray}} 
\newcommand{\X}{\frac{\mathcal{X}}{M_*}}
\newcommand{\Xd}{\frac{\mathcal{X^\dag}}{M_*}}
\newcommand{\R}{\frac{\mathcal{R}}{M_*^2}}
\newcommand{\gR}{{\gamma_\mathcal{R}}}

\title{Phenomenology of SUSY with scalar sequestering
}
\author{%
Gilad Perez$^{a,b}$, Tuhin S. Roy$^c$ and Martin Schmaltz$^{d,e}$\\
\llap{$^a$}
C.~N.~Yang Institute for Theoretical Physics, Stony Brook University,
Stony Brook, NY 11794 \\ 
\llap{$^b$} Department of Particle Physics, Weizmann Institute of
Science, Rehovot 76100, Israel \\
\llap{$^c$} Department of Physics
and Institute of Theoretical Science, \\ 
\hspace*{0.1mm} University of Oregon, Eugene, OR 97403 \\
\llap{$^d$} 
Physics Department, Boston University, 590 Commonwealth Ave, Boston,
MA 02215 \\ 
\llap{$^e$}
Berkeley Center for Theoretical Physics, University of California,
Berkeley, CA 94720  
}
\preprint{WIS/20/08-NOV-DPP}
\abstract{ 
The defining feature of scalar sequestering is that the MSSM squark and slepton
masses as well as all entries of the scalar Higgs mass matrix
vanish at some high scale. This ultraviolet boundary condition - scalar
masses vanish while gaugino and Higgsino masses are unsuppressed - is
independent of the supersymmetry breaking mediation mechanism. It is the
result of renormalization group scaling from approximately conformal
strong dynamics in the hidden sector. We review the mechanism of 
scalar sequestering and prove that the same dynamics which suppresses 
scalar soft masses and the $B_\mu$ term also drives the Higgs soft
masses to $-|\mu|^2$. Thus the supersymmetric contribution to the
Higgs mass matrix 
from the $\mu$ term is exactly canceled by the soft masses.
Scalar sequestering has two tell-tale predictions for the superpartner
spectrum in addition to the usual gaugino mediation predictions:
Higgsinos are much heavier ($\mu \gtrsim$ TeV) than scalar
Higgses ($m_A \sim$ few hundred GeV), and third generation scalar masses
are enhanced because of new positive contributions from Higgs loops.
}

\begin{document}


\section{Introduction}

Tree level sum rules for superpartner masses and the supersymmetric
flavor problem motivate the study of models in which supersymmetry
breaking occurs in a hidden sector and is communicated to the visible
sector through mostly flavor universal interactions. Below the mass
scale of the communication mechanism, the visible and hidden sectors
are only coupled through higher dimensional operators. To compute the
spectrum of superpartner masses in the visible sector one must
renormalize these operators from the communication scale, $M_{*}$,
down to the superpartner mass scale. 

It is by now well-appreciated that hidden sector interactions
can have very large effects on this renormalization. For example, in 
models with conformal sequestering \cite{Luty:2001jh, Luty:2001zv} 
\cite{Dine:2004dv} 
\cite{Ibe:2005pj, Ibe:2005qv, Schmaltz:2006qs} all MSSM superpartner
masses are strongly suppressed relative to the gravitino mass so that
anomaly mediation becomes dominant. In more general models hidden
sector renormalization is not universal, and several new hidden-sector 
dependent parameters are needed to parametrize the superpartner
spectrum~\cite{Cohen:2006qc}  (for recent work in this
direction see~\cite{Kawamura:2008bf, Campbell:2008tt}). When
the hidden sector 
is strongly coupled and approximately conformal between $M_*$ and a
lower scale $M_\text{int}$ then some supersymmetry breaking operators
may be so strongly suppressed that they do not contribute to
superpartner masses. In such cases the number of effective parameters
may actually be reduced. 

For example, references \cite{Roy:2007nz, Murayama:2007ge}
showed that if certain inequalities between anomalous dimensions in
the hidden sector are satisfied then the running scalar masses evolve
to be negligibly small compared to gaugino masses
at $M_\text{int}$\@, and the ultraviolet values of the scalar masses
become irrelevant parameters. We refer to this suppression of scalar
masses  as scalar sequestering. The squark and slepton mass spectrum
of scalar sequestering resembles that of gaugino
mediation~\cite{Kaplan:1999ac, Chacko:1999mi}
\cite{Csaki:2001em,Cheng:2001an}, 
however as was pointed out in  
\cite{Roy:2007nz, Murayama:2007ge} there are important differences
for the Higgs sector. The running Higgs scalar mass matrix in the MSSM 
is 
\begin{equation}
\label{higgs-intro}
  \begin{pmatrix}
    m_{H_u}^2+|\mu|^2 & B_\mu^* \\
    B_\mu & m_{H_d}^2+|\mu|^2 
  \end{pmatrix} \; ,
\end{equation}
where $\mu$ is the usual supersymmetry preserving $\mu$-parameter
while $m_{H_d}^2$,  $m_{H_u}^2$ and $B_\mu$ are supersymmetry
breaking. In the simplest attempts at solving the $\mu$ problem by
coupling the Higgs doublets directly to messenger fields one often
finds that $m_{H_u}^2, m_{H_d}^2,  B_\mu \gg |\mu|^2$\@. This is
problematic~\cite{Dvali:1996cu} because the experimental lower bound
on Higgsino masses 
requires $|\mu| > m_Z$ whereas naturalness of the Higgs potential
gives $m_Z^2 \gtrsim \text{Min}\left[ m_{H_u}^2, m_{H_d}^2,
  B_\mu \right]$. This problem can be solved by scalar
sequestering. We will give a general proof that the same dynamics
which sequesters  the squark and slepton masses also sequesters the
Higgs scalar mass matrix while leaving the $\mu$ term unaffected. 

To summarize, scalar sequestering  predicts a distinct pattern
  of soft parameters at the scale $M_\text{int}$. The non-vanishing 
parameters are
\begin{equation}
\mu, \quad a_t, a_b, a_\tau, \quad M_1, M_2, M_3, 
 \quad m_{H_u}^2 = m_{H_d}^2 = -|\mu|^2  \; ,
\label{boundary}
\end{equation}
where $M_i$ are the gaugino masses and $a_i$ are the A-terms.
Below $M_\text{int}$ the interactions of the hidden sector turn off (by
definition of $M_\text{int}$), and the running is determined by the MSSM
interactions alone.

The boundary values for the soft
masses in Eq.~\eqref{boundary} have several interesting consequences.
Firstly, the fact that the soft scalar masses squared vanish at
$M_\text{int}$ 
ameliorates the supersymmetric flavor problem. Flavor
violation which may have been imprinted 
on the scalar mass operators at higher energies (for example by a 
messenger sector which is not flavor-universal) is rendered harmless
by the sequestering. On the other hand, A-terms are not sequestered by
hidden sector interactions, and large flavor violation in the A-terms
must be avoided. The boundary values in Eq.~\eqref{boundary}
also significantly improve supersymmetric CP problem. 
The fact that $B_\mu$ vanishes at $M_\text{int}$ and our assumption of 
universal gaugino masses allow one to rotate away most of the
flavor-universal CP violating phases. Only the phases of A-terms 
remain and the supersymmetric CP problem is greatly reduced
\cite{Yamaguchi:2002zy}.

Secondly, the fact that the entire scalar Higgs mass matrix vanishes
at $M_\text{int}$ while $\mu$ remains large leads to two unique predictions
for the superpartner spectrum at the TeV scale:
in this model consistent electroweak symmetry breaking
requires $\mu \gtrsim 1$ TeV and therefore heavy Higgsinos.
But at tree level the Higgs scalar masses do not grow with $\mu$ 
and numerically we find $m_A^2 \ll |\mu|^2$ throughout parameter space.
In addition, the negative Higgs soft masses give
positive contributions to the running of the
third generation scalar masses. For example, we expect that the
sum of the masses of the two stau mass eigenstates is larger than
the sum of the selectron or smuon masses.
 
Finally, the fact that there are fewer non-vanishing parameters
at $M_\text{int}$ increases the models' predictivity. For example,
if one also assumes unified gaugino masses and A-terms then the
superpartner spectrum depends on only three free parameters.

The focus of our paper is to determine the superpartner mass spectrum
and phenomenology which follow from the boundary condition in
Eq.~\eqref{boundary}. Section~\ref{sec:model} and the Appendix contain
a review of hidden sector running and the derivation of the predictions
$m_{H_u}^2 = m_{H_d}^2 = -|\mu|^2\;, \  B_\mu =0$, and
$m_Q^2 = m_U^2 = m_D^2 = m_L^2 = m_E^2 = 0$\@. In
Section~\ref{sec:parameter}
we find the viable region of parameter space and derive the spectrum
for a sample point which satisfies all phenomenological constraints. In
Section~\ref{sec:conclusion} 
we conclude and discuss future directions.


\section{A peculiar spectrum at the intermediate scale}
\label{sec:model}

In this Section we review the theoretical framework which leads to the
predicted relations for soft masses shown in
Eq.~\eqref{boundary}. The basic idea is that this pattern of
soft masses is a result of strong renormalization from hidden sector
interactions. This makes it largely independent of the mediation
mechanism operating at high scales.  

We begin by assuming that there is some mediation mechanism between
the visible and hidden sectors which generates a set of higher
dimensional operators coupling the two sectors. The higher dimensional
operators are suppressed by a scale which we denote by $M_*$. For
example, in minimal supergravity, this scale is
the Planck scale. In gauge mediation it is the messenger scale times
$16 \pi^2$\@. At weak coupling, and suppressing indices labeling
different hidden sector operators, the most relevant hidden-visible
couplings are of the form 
\begin{gather}
\int\!\!\! d^4\theta\ \R  Q^\dag Q
        +  \Xd H_u H_d + \R H_u H_d +\ \text{h.c.}
\label{dterms} \\
+ \int\!\!\! d^2\theta\  \X WW + \X Q U^c H_u + \text{h.c.} \ .
\label{fterms}
\end{gather}

Here $Q$ stands for any of the MSSM matter chiral superfields, $W$
stands for the MSSM gauge field strength superfields, $\mathcal{X}$
stands for hidden sector chiral operators, and $\mathcal{R}$ for real
superfield operators of the hidden sector. The hidden sector operators
may be elementary superfields or composite. $\mathcal{R}$ may contain
products of a chiral and an anti-chiral operators
$\mathcal{X^\dagger\! X}$, but in general $\mathcal{R}$ is a sum of
operators, some of which can be written as such products and some
which cannot. The powers of $M_*$ in the denominators have been chosen
according to engineering dimensions so that the operator coefficients
would have no further mass dimensions if $\mathcal{X}$ were a free
chiral superfield $X$ and $\mathcal{R}$ were the product 
$X^\dag X$\@. The real scaling
dimensions of these operators are quite different from the engineering
dimensions and are discussed below. 

Ignoring any renormalization effects for the moment, we obtain the
soft masses of the MSSM by replacing the  hidden sector operators by
vacuum expectation values (VEVs) for their auxiliary components
\begin{equation}
\langle \frac{\left. \mathcal{X} \right|_{F}}{M_*} \rangle
=\frac{F}{M_*} \quad\quad\quad  
\langle \frac{\left. \mathcal{R} \right|_{D}}{M_*^2}
\rangle=\frac{D}{M_*^2} \ .
\label{vevs}
\end{equation}
The couplings in Eq.~\eqref{dterms} become scalar masses, a $\mu$-term
and $B_\mu$ whereas the couplings in Eq.~\eqref{fterms} become gaugino
masses and A-terms, respectively. Without any strong renormalization effects
from the hidden sector a phenomenologically successful model requires
$D\sim F^2$. 

Note that we have omitted any terms of the form $\mathcal{X} Q^\dag Q$
from Eq.~\eqref{dterms} because they can be removed by a field
redefinition. We discuss this field redefinition in more detail in the
Appendix. 

We now turn to the renormalization of the couplings in
Eqs.~\eqref{dterms} and \eqref{fterms} due to hidden sector
interactions.  We work in the holomorphic basis for hidden sector
fields so that supersymmetric non-renormalization theorems are manifest. 

An operator which is chiral or anti-chiral in hidden sector fields
(i.e. an operator which depends on $\mathcal{X}$ or $\mathcal{X}^\dag$
only) is not renormalized by purely hidden sector interactions in
the holomorphic basis for hidden sector fields (see \cite{Roy:2007nz} 
for a proof). This immediately implies that the operators for
gaugino masses and A-terms in Eq.~\eqref{fterms}
and the $\mu$ term in Eq.~\eqref{dterms} are not
renormalized by hidden sector interactions. This
non-renormalization theorem extends to the actual gaugino masses,
$A$-terms and and the $\mu$-term if they are expressed in terms of the
expectation value for the holomorphic operator $\mathcal{X}$\@.

The operators which involve $\mathcal{R}$ are not protected from
renormalization. They receive anomalous dimensions from hidden sector
interactions 
which can have either sign \cite{Roy:2007nz} and are not calculable at
strong coupling. The crucial dynamical assumption that underlies our
framework is that these anomalous dimensions are large and positive
\cite{Roy:2007nz, Murayama:2007ge, Giudice:2007ca}. 
Then all operators involving $\mathcal{R}$
are strongly suppressed at low energies. More precisely, we assume
that the hidden sector is governed by a strongly coupled approximate
fixed point below the scale $M_*$ and down to the scale
$M_\text{int}$. Any operator in the low energy effective Lagrangian
involving 
$\mathcal{R}$ will then be suppressed by a factor of
$(M_\text{int}/M_*)^{\gR}$ where $\gR$ is the anomalous dimension of
the operator $\mathcal{R}$. 
When $\gR$ is of order one and the range of scales over which the
strong hidden sector interactions operate is large, then all operators 
involving $\mathcal{R}$ have small coefficients at $M_\text{int}$, and
their contributions to the running superpartner masses at
$M_\text{int}$ can be neglected.  This is nice for two reasons: 

\begin{itemize}
\item The operators of the form $\R Q^\dag Q$ may have non-trivial
  flavor structure from flavor physics in the ultraviolet.
  The resulting mass matrices for squarks and
  sleptons violate flavor and lead to flavor changing neutral currents
  which are tightly constrained by experiment. 
  Hidden sector running suppresses such flavor violation and
  might therefore make some flavor-violating mediation mechanisms viable.

\item Electroweak symmetry breaking requires that the coefficient of
  the operator $\R H_u H_d$ which gives rise to the $B_\mu$ term after
  supersymmetry breaking is small. More specifically, one needs $B_\mu
  \sim M_\text{susy}^2/\tan \beta$ in the infrared. A small
  coefficient for   $\R H_u H_d$ is exactly what our renormalization
  factor   predicts. Note however that our mechanism predicts small
  $B_\mu$ at $M_\text{int}$, whereas electroweak symmetry breaking
  requires small $B_\mu$ at the TeV scale. Therefore MSSM running
  below $M_\text{int}$ should not
  generate very large contributions to $B_\mu$. This will play a
  significant role in Sec.~\ref{sec:parameter}.  
\end{itemize}

\noindent Given our assumptions about the anomalous dimensions of $\R$
a very simple and attractive picture emerges: at the scale
$M_\text{int}$ all operators involving $\R$ are suppressed and can be
neglected. The soft terms are then determined by the remaining
operators    
\begin{equation}
\int\!\!\! d^4\theta\ \Xd H_u H_d +\int\!\!\! d^2\theta\  \Big[\;
\X Q U^c H_u
+ \X WW \Big] + \text{h.c.} \ .
\label{relterms}
\end{equation}
which give rise to the $\mu$ term, A-terms, and gaugino masses at the
scale $M_\text{int}$, respectively. TeV scale parameters are
computed by using the usual MSSM renormalization group equations to
evolve from $M_\text{int}$ down to the weak scale. By assumption
hidden sector interactions are not strongly coupled below the scale
$M_\text{int}$, therefore they do not contribute significantly to this
running. Thus the entire spectrum of soft masses is given in terms of
only a few parameters: 
\begin{equation}
\mu, \quad a_t, a_b, a_\tau, \quad M_1, M_2, M_3, \quad
\log\frac{M_\text{int}}{M_\text{susy}}, \quad \lambda_t \, . 
\end{equation}
Many mediation mechanisms predict gaugino mass unification, we therefore
assume $M_1/g_1^2=M_2/g_2^2=M_3/g_3^2 \equiv M_u$.
 For simplicity we also assume universal A-terms, 
$a_t/\lambda_t = a_b/\lambda_b = a_\tau/ \lambda_\tau =  A_u $, 
however for the relatively small values of $\tan \beta$  which we consider
the contributions from $a_b$ and $a_\tau$ to the superpartner spectrum
are not very significant. 

The remaining free parameters are then 
\begin{equation}
\mu, \quad A_u, \quad M_u, \quad
\log\frac{M_\text{int}}{M_\text{susy}}, 
\quad \lambda_t \,, 
\end{equation}
These parameters (together with the gauge couplings) determine the
Higgs potential which in turn determines the electroweak symmetry
breaking VEVs $v_u$ and $v_d$, or equivalently $v=246$ GeV and $\tan
\beta$. Fitting to the measured top and W masses fixes the top
Yukawa, $\lambda_t$, and one of the mass parameters, leaving a
3-dimensional parameter space. In the next Section we will explore
this parameter space. We will see that there are choices for the
parameters which avoid all experimental constraints, but that the
requirements of consistent electroweak symmetry breaking and lower
bounds on particle masses are enough to tightly constrain the allowed
region in parameter space.

Before closing this Section we must discuss an important subtlety in
the renormalization of the Higgs soft masses due to hidden sector
interactions. This leads to an interesting
modification of the boundary conditions at the scale
$M_\text{int}$. The subtlety is that the $\mu$-term operator $\Xd H_u
H_d$ contributes to the renormalization of the Higgs soft mass
operators $\R H_u^\dag H_u$ and $\R H_d^\dag H_d$ \cite{Roy:2007nz}. 
One can
prove that the soft masses of the Higgses do not run to zero like all
the other soft scalar masses. Instead they run to a quasi-fixed point
which predicts  
\begin{equation}
m_{H_u}^2=m_{H_d}^2=-|\mu|^2
\label{softhiggsmass}
\end{equation}
at $M_\text{int}$\@. We give a proof for this equation in the
Appendix. In summary, our model is defined by the following
boundary conditions at the scale $M_\text{int}$\@: 
\begin{align}  
  \frac{a_t}{\lambda_t}=\frac{a_b}{\lambda_b}=\frac{a_\tau}{\lambda_\tau}\equiv
  A_u, \quad \quad   
       \frac{M_1}{g_1^2}= \frac{M_2}{g_2^2}= \frac{M_3}{g_3^2} \equiv
        M_u, \nonumber \\ \nonumber \\ 
 m^2_{H_u}=m^2_{H_d}=-|\mu|^2, \quad m^2_{Q,U,D,L,E}=0, \quad B_\mu=0 \; .
\label{spectrum-int}
\end{align}

In Section 3 we will explore electroweak symmetry breaking with this
boundary condition and find the superpartner spectrum for a
representative point in parameters space.

\section{Electroweak symmetry breaking and a sample spectrum}
\label{sec:parameter}

In the previous section we derived boundary conditions for the soft
supersymmetry breaking parameters of the MSSM at the intermediate
scale. We found that the entire superpartner mass spectrum depends on
only 5 free parameters. Two combinations of these parameters can be
fixed by demanding that our model correctly reproduce the measured top
and $Z$ masses. One of our goals in this section is to map out the
remaining three-dimensional  parameter space. The conditions for
radiative electroweak symmetry breaking and stability of the vacuum
significantly constrain parameter space. In particular, We find that
the intermediate scale must be fairly high and that $\mu$ is on the
order of the gluino mass.

In the previous Section we have seen that the superpartner spectrum
depends on the following five parameters 
\begin{equation}
\lambda_t, \qquad  \log\frac{M_\text{int}}{M_\text{susy}}, \qquad \mu,
\qquad  M_0,   \qquad A_u \;  .
\label{5inputs}
\end{equation}
The two conditions which ensure that we reproduce the correct $Z$ and
top masses at the electroweak scale can be written as  
\begin{align}
  \frac{| m_{H_u}^2 - m_{H_d}^2|}{\sqrt{ 1 - \sin^2\beta}} -
        \left( m_{H_u}^2 + m_{H_d}^2 + 2|\mu|^2 \right) &= m_Z^2 \;
\label{eq:ewsb1}  \\
    B_\mu  -
    \frac{\tan\beta}{1+\tan^2\beta} \left(  m_{H_u}^2 + m_{H_d}^2
    + 2|\mu|^2   \right) &= 0  \;
\label{eq:ewsb2}
\end{align}

Two among the five parameters in Eq.~\eqref{5inputs}  can be
eliminated using the two equations in Eqs.~\eqref{eq:ewsb1} and
\eqref{eq:ewsb2}, but the selection of which parameters to eliminate
is arbitrary. In phenomenological studies of the MSSM usually $\mu$
and $B_\mu$ are solved for as functions of the other parameters. This
is possible and convenient in models where $\mu$ and $B_\mu$ are free
parameters because they do not enter the renormalization group
equations of any other parameters. Therefore $\mu$ and $B_\mu$ can
simply be determined at low energies from Eqs.~\eqref{eq:ewsb1} and
\eqref{eq:ewsb2}.  However, this strategy does not work here because
$B_\mu$ is not a free parameter and because the value of $\mu$ enters
the renormalization of several other soft masses through the initial
conditions $m_{H_u}^2= m_{H_d}^2 =- |\mu|^2$ at the intermediate scale.  

Instead, we will find it convenient to choose $y_t$ (or equivalently
$\tan \beta$) and the two dimensionless parameters $\log
M_\text{int}/M_\text{susy}$, $\hat \mu \equiv \mu/M_u$ as inputs. We
then use Eqs.~\eqref{eq:ewsb1} and \eqref{eq:ewsb2} to solve for
$\hat A \equiv A_u/M_u$ and the overall mass scale of soft masses $M_u$ 
\begin{equation}
\begin{split}
\label{eq:free-param}
  \text{input:}&   \qquad              
  \tan  \beta, \quad \log \frac{M_\text{int}}{M_\text{susy}}, \quad
  \hat \mu = \frac{\mu}{M_u}  \\ 
  \text{solved for:}&  \qquad 
         \hat A = \frac{A_u}{M_u}, \quad M_u \;  .
\end{split}
\end{equation} 
More specifically, we factor out the overall mass scale $M_u$ from
Eq.~\eqref{eq:ewsb2} and then find $\hat A$ as a function of the other
parameters. Then we use Eq.~\eqref{eq:ewsb2} to determine $M_u$.

\begin{figure}[h]
\centering
\begin{minipage}[c]{0.45\linewidth}
   \centering
   \includegraphics[width = 2.95 in]{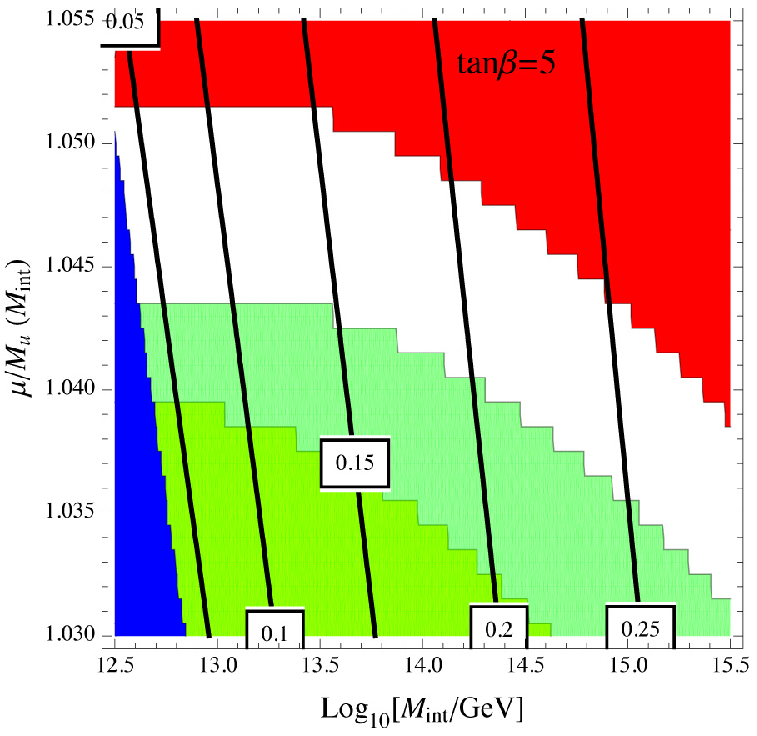} 
\end{minipage}
\hspace{0.5 cm}
\begin{minipage}[c]{0.45\linewidth}
   \centering
   \includegraphics[width = 2.95 in]{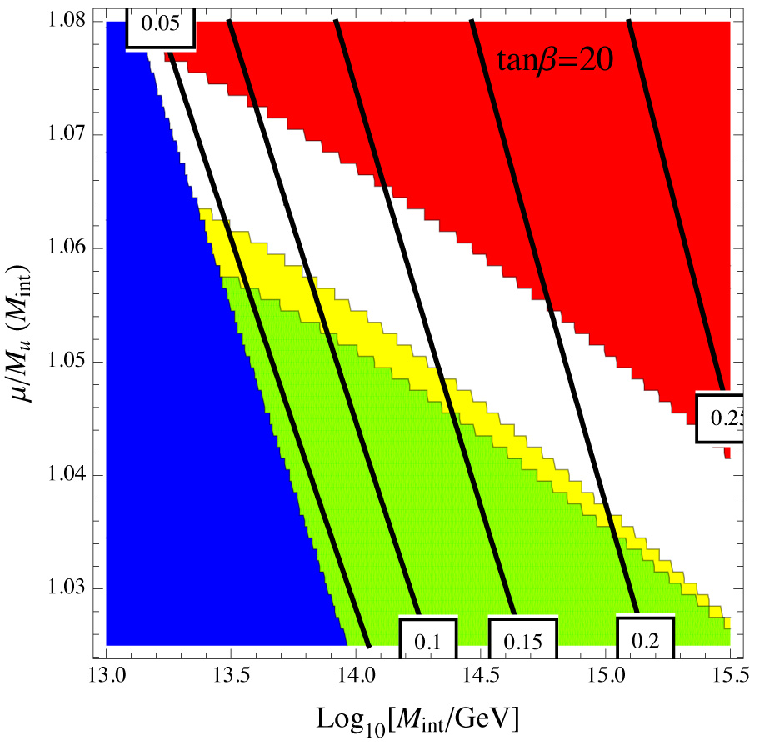}
\end{minipage}  
  \caption{The Figures on the left and right show allowed regions
(white) in the 
$\left[\hat \mu - \log (M_\text{int}/\text{GeV})\right]$ plane for 
$\tan \beta=5$ and $20$ respectively. In the red region the Higgs soft
mass matrix does not have a negative eigenvalue and the vacuum is
unstable in the blue region. 
The Higgs is lighter than 114
GeV in the green region, and the 
right-handed sleptons are lighter than the LEP bounds in the yellow region.
In the yellow-green region both the Higgs and a slepton are lighter than
their LEP bounds. The Higgs masses for these figures were computed in the
decoupling limit~\cite{Martin:1997ns}. Finally, the black contours
lines are contours of $m_A/\mu$\@.} 
 \label{fig:1}
\end{figure}

To visualize the allowed parameter space we choose two representative
values for $\tan\beta$, $\tan \beta = 5 \text{ and } 20$, and plot the
parameter 
space as a function of the other two inputs, $\log
(M_\text{int}/M_\text{susy})$ and $\hat \mu$ in Fig.~\ref{fig:1}.
The allowed region is bounded by a number of constraints:
{\it i.} one eigenvalue of the soft Higgs mass matrix must be negative
so that electroweak symmetry breaking is triggered, $  \left( m_{H_u}^2
  + |\mu|^2 \right) \left( m_{H_d}^2 + |\mu|^2 \right)- |B_\mu|^2 \; <
\; 0 $, {\it ii.} vacuum stability requires $m_{H_u}^2 + m_{H_d}^2 +
2|\mu|^2 \; > \; 2\: |B_\mu| $, {\it iii.} the Higgs mass bound from
LEP $m_h > 114$~GeV, {\it iv.} the bound on the mass of the
right-handed sleptons $m_{\twi e, \twi \mu} > 100$~GeV and  $m_{\twi
\tau} > 91$~GeV.   

The remaining region in parameter space is quite
constrained. Everywhere in parameter space we find $\hat \mu\gtrsim 1$
or $\mu\gtrsim M_u$. Furthermore we see that the intermediate scale is
required to be quite high. This can be understood by looking at the
renormalization group equation for the trace of the $H_u$-$H_d$ mass
matrix which needs to be positive for vacuum stability but is driven
negative by stop loops for small $M_\text{int}$. The plots we show
here are for positive $\mu$, a similar region in parameter space is
allowed for negative $\mu$. 

As mentioned above $\mu$ is required to be quite large in order to
obtain consistent electroweak symmetry breaking. This implies heavy
Higgsinos. Usually in the MSSM large $\mu$ also implies large scalar
Higgs masses and therefore large $m_A$. This is not the case in our
scenario because of the cancellation between the soft mass $-|\mu|^2$
and supersymmetric mass $+|\mu|^2$ in the Higgs mass matrix.  At the
loop level,{\it  i.e.} including running, the Higgs scalar mass matrix
does not vanish but Higgs scalar masses remain much smaller than
Higgsino masses. Hence we expect that the ratio of the pseudoscalar
Higgs mass $m_A$ over Higgsino masses $\mu$ is small. This expectation
is borne out by our numerical analysis as can be seen from the
$m_A/\mu$ contours in Fig.~\ref{fig:1}.

\begin{table}[h]
  \centering
  \begin{tabular}[c]{ | c || c | c |  c | c |}
  \hline
         & \multicolumn{2}{|c|} {$\tan \beta$} & 
         \multicolumn{2}{|c|} {$10$}   \\ 
  Input  & \multicolumn{2}{|c|} {$\hat \mu $ }   &  \multicolumn{2}{|c|} {$1.00$ }  \\
         & \multicolumn{2}{|c|} {$M_\text{int}$} &
           \multicolumn{2}{|c|} {$ 10^{15}$~GeV}  \\
  \hline
  \hline
         & $ M_1$ & $273$~GeV &  $ m_{Q_1}$ & $1243$~GeV \\  
         & $ M_2$ & $510$~GeV &  $ m_{u_1}$    & $1192$~GeV \\
         & $ M_3$ & $1412$~GeV &  $ m_{d_1}$ & $1186$~GeV  \\
         & $ \mu$ & $1246$~GeV  &  $ m_{t_1}$ & $1113$~GeV \\
         & $ B_\mu$ & $ \left(115~\text{GeV} \right)^2 $
              &  $ m_{t_2}$ & $1277$~GeV \\
  Output & $ m_h$ & $115$~GeV  &  $ m_{b_1}$ & $1279$~GeV  \\
         & $m_A$  & $365$~GeV &  $ m_{b_2}$ & $1226$~GeV \\
         & $m_{H^0}$ &  $377$~GeV & $ m_{L_1}$ & $389$~GeV  \\
         & $m_{H^\pm}$ &  $374$~GeV & $ m_{E_1}$ & $204$~GeV  \\
         & $a_t$      & $-906$~GeV &  $ m_{\tau_1}$ & $206$~GeV  \\
         &   &  &  $ m_{\tau_2}$ & $397$~GeV  \\
   \hline
  \end{tabular}
  \label{table:spectra}
  \caption{A sample spectrum.  We evaluated all soft masses at
    $M_\text{susy} = 1~\text{TeV}$.
    Gauge couplings at $M_\text{susy}$ are derived from their $Z$-pole
    values using one-loop standard model RGEs. Yukawa couplings at
    $M_\text{susy}$ are derived from the running quark masses in the
    standard model at $1$~TeV~\cite{Xing:2007fb}. All masses shown in
    the table are determined from the soft masses at one-loop level 
    via tree level matching, except for the lightest neutral Higgs
    mass for which we included higher order
    corrections~\cite{Djouadi:2002ze}. The MSSM soft parameters were
    run at one loop using \cite{Martin:1997ns}, all soft masses are
    evaluated at $M_\text{susy}= 1$~TeV. }   
\end{table}

We believe that this prediction, $m_A^2 \ll |\mu|^2$, is unique to our
scenario. A tell-tale signature which distinguishes our model from
otherwise similar gaugino mediation models is therefore that charged
Higgses $H^\pm$, the pseudoscalar Higgs $A^0$, and the heavy Higgs
$H^0$ can all be produced directly or in cascade decays of stops and
sbottoms with large cross sections at the LHC. On the other hand,
Higgsinos are too heavy to be produced either directly or in cascades.

Another consequence of the negative Higgs soft masses at
$M_\text{int}$ is that third generation scalar masses receive
additional positive contributions from the running due to scalar Higgs
loops. Consider for example the renormalization group equations
for the soft masses squareds of staus. They contain a term proportional
to $\lambda_\tau^2(m_{L_3}^2+m_{E_3}^2+m_{H_d}^2)$. Since $M_{H_d}^2$ is
negative and large this gives a positive contribution to the stau masses
which is absent for smuons and selectrons because of the much smaller
Yukawa couplings. One might still end up with a stau being the
lightest slepton if there is large mixing between left- and right-handed
staus. However the combination $m_{\tau_1}^2+m_{\tau_2}^2$
is independent of this mixing, and we predict that this is
greater than $m_{L_i}^2+m_{E_i}^2$ for $i\in\{1,2\}$. 
Note that this predicted inequality is satisfied by the our example
spectrum in Table~\ref{table:spectra}. 
A similar argument applies to squarks. There the
relevant combination $\lambda_t^2(m_{Q_3}^2+m_{U_3}^2+m_{H_u}^2)$
starts out negative
near $M_\text{int}$, however it turns positive because of the 
large squark masses generated from gaugino loops. Therefore we do
not predict $m_{t_1}^2+m_{t_2}^2$ to be larger than the corresponding
first and second generation squark masses squared. However we do expect
$m_{t_1}^2+m_{t_2}^2$ larger than in usual MSSM spectra in which the
contributions from negative Higgs soft masses are absent. For example, for
the spectrum of Table~\ref{table:spectra} we have
$m_{t_1}^2+m_{t_2}^2= (1694~\text{ GeV})^2$ 
to be compared with $m_{t_1}^2+m_{t_2}^2=(1535~\text{ GeV})^2$ which was
obtained with the same boundary conditions at $M_\text{int}$ except that
we set $m_{H_u}^2=m_{H_d}^2=0$.

We ran the MSSM masses at one loop using \cite{Martin:1997ns}.
The masses quoted in Table.~\ref{table:spectra}  are evaluated at
$M_\text{susy}= 1$~TeV.

Other features of our spectrum are shared with gaugino 
mediation~\cite{Schmaltz:2000ei}. For example, 
right-handed (charged) sleptons are the lightest MSSM
superpartners~\cite{Dimopoulos:1996vz, Dimopoulos:1996fj}. 
Obviously these cannot be the dark matter and the
gravitino or an axion or another particle in addition to the MSSM may
be the dark matter. 
Cascade decays in our model always end in right-handed sleptons. Depending
on their lifetime, these may manifest themselves either as stable charged
tracks or as displaced vertices from their decays to leptons and
gravitinos~\cite{Fayet:1977vd,Fayet:1979sa,Casalbuoni:1988kv}.

\section{Conclusions}
\label{sec:conclusion}
In this paper we have shown that the scalar sequestering boundary
condition with gaugino mass unification is compatible with radiative 
electroweak symmetry breaking, and that it produces a viable superpartner 
spectrum. Scalar sequestering drives all scalar masses to zero at an
intermediate scale. Therefore any dependence of scalar masses on details of
the messenger sector of supersymmetry breaking is removed by
renormalization, and we find several messenger-model independent
predictions for the superpartner spectrum. 
The running squark and slepton masses pass through zero
at the intermediate scale when evolved with the MSSM renormalization
group equations as in gaugino mediation. In addition, there are unique
predictions which follow from the vanishing Higgs scalar mass matrix
\begin{equation}
  \begin{pmatrix}
    m_{H_u}^2+|\mu|^2 & B_\mu^* \\
    B_\mu & m_{H_d}^2+|\mu|^2 
  \end{pmatrix} =
  \begin{pmatrix} 0&0 \\ 0&0 \end{pmatrix} \; .
\end{equation}
These predictions are: Higgs scalars are much lighter than Higgsinos.
Third generation sleptons (and to some extent also squarks, see the
discussion near Table 3. for details) are lighter than their first and
second generation counterparts due to the contributions from negative
Higgs soft masses in the renormalization group equations.  

We find the vanishing of the Higgs scalar mass matrix at $M_\text{int}$
also very intriguing from a theoretical point of view. It is
interesting that supersymmetry breaking parameters $m_{H_u}^2$ and
$m_{H_d}^2$ become 
related to the supersymmetry preserving parameter $\mu$ by hidden sector
running. This is at least a partial solution to the $\mu$-problem once
one realizes that the $\mu$ problem may be formulated as the need for
an explanation for why the combinations $m_{H_u}^2+|\mu|^2$ and
$m_{H_d}^2+|\mu|^2$ are small compared to $m_{H_u}^2$, $m_{H_d}^2$ and
$|\mu^2|$ individually. 

In this paper we minimized the number of free parameters by making the
additional assumption of gaugino mass unification. We found that the
allowed region in parameter space is very small, leading to very
specific 
predictions for the spectrum. 
Clearly, if one allows the gaugino masses to vary independently, it
becomes much easier to find solutions to the electroweak symmetry
breaking conditions. We believe that such models might have very low
levels of fine tuning in the Higgs sector.  

Finally, we wish to comment on the recent paper
Ref.~\cite{Asano:2008qc} which has some overlap with our work. The
authors of Ref.~\cite{Asano:2008qc} assumed the  relationship $|\mu^2|
=|a_t a_b|$ which  follows from specific assumptions about the
couplings of $H_u$ and $H_d$ to the messenger sector. 
With this additional constraint they found no solutions to the
electroweak symmetry breaking conditions Eqs.~\eqref{eq:ewsb1} and
\eqref{eq:ewsb2}. 
To avoid this problem, the authors of \cite{Asano:2008qc} introduced
additional free parameters for the gaugino masses.
In our paper we did not make any assumptions about the relationship
between $\mu$ and the $A$-terms because such relationships are
messenger-model dependent. We found that then there are solutions to
the electroweak symmetry breaking conditions without the need to give
up gaugino mass unification.

\begin{acknowledgments}
We thank V.~Sanz for collaboration in the early stages of this work and 
and G.~Kribs for stimulating discussions. GP thanks Boston University and
Harvard University for their hospitality. MS thanks the Center for
Theoretical Physics at Berkeley and the LBNL particle theory group for
their hospitality. This work was supported in part by an NSF grant
PHY-06353354 (GP) and the DOE under contracts DE-FG02-96ER40969 (TSR),
DE-FG02-91ER-40676 (MS) and  DE-FG02-01ER-40676 (MS).

\end{acknowledgments}

\appendix

\section{The proof}
\label{appendixproof}

We wish to prove Eq.~\eqref{softhiggsmass}, i.e. that the coefficients
of the Higgs soft mass operators run to fixed point values equal to
minus the square of the coefficient of the $\mu$-term operators. Our
context is the MSSM coupled to an approximately conformal hidden
sector through higher-dimensional operators. We also assume that the hidden
sector couplings are sufficiently large so that their effects dominate
over any running due to MSSM couplings, and it is a good approximation
to ignore the MSSM couplings.%
\footnote{The size of corrections due to non-zero MSSM couplings, $g$,
  is straightforward to estimate. They are proportional to $g^2/(16
  \pi^2 \gR)\sim 10^{-2}$ where $\gR$ is the anomalous dimension
  defined in the text.}

As a warm-up let us first discuss the renormalization of the Higgs
soft masses in absence of the $\mu$ and the $B_\mu$ 
term operators but with the
most general coupling between the hidden and visible sectors. Since we
are ignoring visible sector interactions it suffices to look at a
single chiral superfield of the visible sector $H$ coupled to hidden
sector operators $\mathcal{X}$ and $\mathcal{R}$ 
\begin{equation}
\int\!\!\! d^4 \theta \ H^\dag\! H \: (1+x \, \X+ x^\dag \, \Xd +
x^\dag x \, \frac{\mathcal{X^\dag X}}{M_*^2}+ r \R) 
     + \mathcal{L}_\text{hidden} \; .
\label{appgeneralL}
\end{equation}
We assumed that operators with derivatives are less relevant so that
we can ignore them.  
The chiral operator $\mathcal{X}$ may actually consist of a sum of
terms so that $x$ is a complex vector of coefficients. The operator
$\mathcal{R}$ is real and the corresponding coefficient vector $r$ is 
real as well. Note that we could have absorbed  the term proportional
to $x^\dag x$ by an appropriate shift in $r$ but we will see that
separating the operators in this way is preferable. 

The easiest way to understand the renormalization of the scalar mass
operators is to first redefine fields to remove the chiral couplings
of $\mathcal{X}$ to $H^\dag H$. To do so we define 
\begin{equation}
\tilde H \equiv (1 + x \, \X) H \; .
\label{appfieldredef}
\end{equation}
Ignoring operators of order $(\X)^3$ or higher our Lagrangian becomes
\begin{equation}
\int\!\!\! d^4\theta\ \tilde H^\dag \! \tilde H \: (1 + r \, \R) +
\mathcal{L}_\text{hidden} \ , 
\end{equation}
and all dependence on $x$ has disappeared, this was the reason for
splitting out the $x^\dag x$ term from $r$ in
Eq.~\eqref{appgeneralL}. Since the Lagrangian is independent of $x$ it
is clear that the running of operator coefficients $r$  is independent
of $x$ and $\mathcal{X}$, and also the scalar masses are independent
of $x$ and $\mathcal{X}$\@.  

We now assume that our hidden sector is a strongly coupled
approximately conformal field theory so that the running of operator
coefficients can be approximated by anomalous dimensions. We further
assume that the anomalous dimensions $\gR$ of the real operators
$\mathcal{R}$ are positive. This is a strong assumption but given
the large number of approximately conformal field theories which we
can construct we believe that it is a reasonable assumption that
hidden sectors with the desired properties exist~\cite{Roy:2007nz}.%
\footnote{We do not know of any techniques for computing all the
  anomalous dimensions of real operators $\mathcal{R}$ in strongly
  coupled $N=1$ supersymmetric field theories. A weakly coupled
  example in which we can compute the anomalous dimensions in
  perturbation theory is the theory of a single chiral superfield $X$
  with the superpotential coupling $W=\lambda X^3/3!$. In this
  example, the anomalous dimension of $\mathcal{R}=X^\dag \! X$ is
  positive and $\gR = +2 \lambda^2/16 \pi^2$\@.} 
The renormalization group equation for the coefficients $r$ is then
(the Feynman diagrams which contribute are of the
form of the third diagram in Figure 2.)
\begin{equation}
\frac{d\, r}{dt}=\gR \, r
\end{equation}
with the low energy solution
\begin{equation}
r \,\Big|_{M_\text{int}}\: = \: 
    \left(\frac{M_\text{int}}{M_*}\right)^\gR  \: r \, \Big|_{M_*}\ .
\end{equation}
Thus the resulting scalar masses squared are suppressed by the factor
$(M_\text{int}/M_*)^{\gR}$ relative to the gaugino masses which have
no such suppression. {\it i.e.} the scalar masses are sequestered.

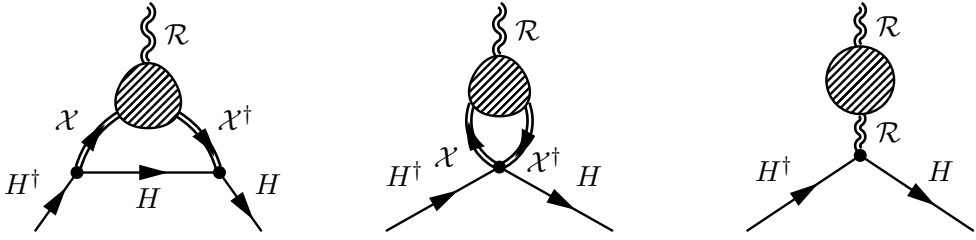
\begin{figure}[t]
\centering
\setlength{\unitlength}{1.5 mm}
\begin{fmffile}{figure1}
  {
    \begin{fmfgraph*}(20,20) 
      \fmfstraight
      \fmfbottom{b1,b2}\fmftop{t1}
      \fmfpolyn{shaded,smooth,pull=1.8,tension=0.8}{H}{3}
      \fmf{phantom}{b1,v1,H1}
      \fmf{phantom}{H2,v2,b2} 
      \fmf{phantom,tension=2}{t1,H3}  \fmffreeze
      \fmf{double,label=$\mathcal{X}$,l.side=left,left=0.2}{v1,H1}  
      \fmf{double,label=$\mathcal{X}^\dag$,l.side=left,left=0.2}{H2,v2} 
      \fmf{dbl_wiggly,label=$\mathcal{R}$,tension=2}{H3,t1} 
      \fmf{phantom_arrow,left=0.2}{v1,H1}   
      \fmf{phantom_arrow,left=0.2}{H2,v2}   
      \fmf{fermion,label=$H^\dag$,l.side=left}{b1,v1}
      \fmf{fermion,label=$H$,l.side=right}{v1,v2}
      \fmf{fermion,label=$H$,l.side=left}{v2,b2}
      \fmfdotn{v}{2}
    \end{fmfgraph*}
  }\hfil
  {
    \begin{fmfgraph*}(20,20)   
       \fmfstraight
      \fmfbottom{b1,b2}\fmftop{t1}
      \fmfpolyn{shaded,smooth,pull=1.8,tension=0.5}{H}{3}
      \fmf{phantom}{b1,v1,H1}
      \fmf{phantom}{H2,v1,b2} 
      \fmf{phantom,tension=2.5}{t1,H3}  \fmffreeze
      \fmf{double,label= $\mathcal{X}$,l.side=left,left=0.4}{v1,H1}  
      \fmf{double,label=$\!\!\!\mathcal{X}^\dag$,l.side=left,left=0.4}{H2,v1} 
      \fmf{dbl_wiggly,label=$\mathcal{R}$,tension=2}{H3,t1} 
      \fmf{phantom_arrow,left=0.4}{v1,H1}   
      \fmf{phantom_arrow,left=0.4}{H2,v1}   
      \fmf{fermion,label=$H^\dag$ ,l.side=left}{b1,v1}
       \fmf{fermion,label=$\ H$,l.side=left}{v1,b2}
       \fmfdotn{v}{1} 
    \end{fmfgraph*}
  } \hfil
  {
    \begin{fmfgraph*}(20,20)   
      \fmfstraight
      \fmfbottom{i1,o1} \fmftop{t1}
      \fmf{fermion,label=$H^\dag$,l.side=left}{i1,v1}
      \fmf{fermion,label=$H$,l.side=left}{v1,o1}
      \fmf{dbl_wiggly,label=$\mathcal{R}$,l.side=left,tension=2}{t1,v2} 
      \fmf{dbl_wiggly,label=$\mathcal{R}$,l.side=left,tension=2}{v2,v1}      
      \fmfdotn{v}{2}  \fmfblob{.3w}{v2}
    \end{fmfgraph*}
  } 
\end{fmffile}
\caption{Renormalization of the Higgs soft mass operators 
$\mathcal{R} H^\dag H$ due to the operators 
$\left(\mathcal{X}+\mathcal{X}^\dag\right) H^\dag H$\@,
$\mathcal{X^\dag X} H^\dag H$ and $\mathcal{R} H^\dag H$ itself.}
\label{fig:dia1}
\end{figure}

Note that we could have derived the same result without performing the
field redefinition of Eq.~\eqref{appfieldredef}. In this basis the
renormalization is slightly more complicated because the operators $\X 
H^\dag H$ and $\frac{\mathcal{X^\dag \! X}}{M_*^2}H^\dag H$ now also
contribute to the renormalization of $r$ via the two left-most diagrams in
Fig.~\ref{fig:dia1}. Note that the ``blobs'' in these two diagrams
represent identical hidden sector interactions. 
Therefore the diagrams are proportional to identical unknown hidden
sector factors. But the first diagram is proportional to $(-ix)^\dag
i(-ix)=+i x^\dag x$ from the two vertices and the ``propagator'' for
the F-component of $H$, whereas the second diagram has a $-i x^\dag x$
from the vertex. Therefore the two diagrams cancel, and the renormalization
group equation for $r$ only receives contributions from the third diagram
in Fig.~\ref{fig:dia1}.
Thus we see that the running of $r$ is the same as in the other
basis, and $r$ is suppressed by  $(M_\text{int}/M_*)^\gR$ in the
infrared. Note 
that the $x$-dependent part of the Lagrangian does not run (the term
linear in $x$ is protected by the non-renormalization theorem, and the
coefficient $x^\dag x$ is equal to the square of the linear $x$
coefficient by definition). 
At $M_\text{int}$ the $r$ terms can be neglected, and the
remaining  Lagrangian is  
\begin{equation}
\int \!\!\! d^4\theta\ H^\dag \!H \: 
 \left(1+ x \, \X+ x^\dag \, \Xd + x^\dag x \,    
     \frac{\mathcal{X^\dag \! X}}{M_*^2} \right) 
  + \: \mathcal{L}_\text{hidden} \ . 
\end{equation}
This Lagrangian does not give $H$ scalar masses because 
the contributions from $F$-terms
in $\X$ cancel the masses from
$D$-terms in $\frac{\mathcal{X^\dag \! X}}{M_*^2}$. This can be seen 
explicitly by integrating out the auxiliary components of $H$ or -
more easily - by performing the redefinition to $\tilde H$ fields. 

This completes our study of the renormalization of the Higgs soft mass in
the presence of the coupling $\frac{\mathcal{X}}{M_*} H^\dag H$, but
without the operators responsible for generating the $\mu$ and $B_\mu$ terms.
The reason for considering this simpler case first is that the running of the Higgs
soft mass operators ($\frac{\mathcal{X}^\dag \mathcal{X}}{M_*^2} H^\dag
H$) due to the $\mu$-term operator can be understood by performing a
similar field redefinition to the one given in
Eq.~\eqref{appfieldredef}. 
We now turn to our
proof in the general case which includes the $\mu$-operator 
$\frac{\mathcal{X}^\dag}{M_*} H_u H_d$\@.

Ignoring any (weak) visible sector interactions the relevant
Lagrangian which couples $H_u$, $H_d$ to the hidden sector fields is 
\begin{equation}
\int\!\!\! d^4\theta\ \Bigg[ H_u^\dag  H_u + H_d^\dag  H_d 
  + x_\mu^\dag \: \X H_d^\dag  H_u^\dag + x_\mu \! \Xd H_u H_d 
+ r_u \: \R H_u^\dag  H_u + r_d \: \R H_d^\dag  H_d \Bigg] \ .
\label{appmuL}
\end{equation}
Note that we have written the Lagrangian directly in the basis for
$H_u$ and $H_d$ in which there are no chiral $\mathcal{X}$ couplings
to $H_u^\dag H_u$ and $H_d^\dag H_d$\@.%
\footnote{The field redefinition required for going to this basis
  generates $A$-terms proportional to Yukawa couplings and also a
  $B_\mu$ -term proportional to the $\mu$-term. These terms are not
  relevant to the hidden sector induced renormalization of the scalar
  masses.} 
There is also the operator $\mathcal{R} H_u H_d$ which contributes to
$B_\mu$\@. This operator scales to zero in the infrared because of the
anomalous dimension of $\mathcal{R}$ which explains why $B_\mu \simeq
0$ at $M_\text{int}$\@. Since $\mathcal{R} H_u H_d$ does not contribute
to the renormalization of the soft masses $m_{H_u}^2$ and $m_{H_d}^2$
we have not included it in Eq.~\eqref{appmuL}.

\begin{figure}[t]
\centering
\setlength{\unitlength}{1.5 mm}
\begin{fmffile}{figure2}
  { 
    \begin{fmfgraph*}(20,20) 
      \fmfstraight
      \fmfbottom{b1,b2}\fmftop{t1}
      \fmfpolyn{shaded,smooth,pull=1.8,tension=0.8}{H}{3}
      \fmf{phantom}{b1,v1,H1}
      \fmf{phantom}{H2,v2,b2} 
      \fmf{phantom,tension=2}{t1,H3}  \fmffreeze
      \fmf{double,label=$\mathcal{X}$,l.side=left,left=0.2}{v1,H1}  
      \fmf{double,label=$\mathcal{X}^\dag$,l.side=left,left=0.2}{H2,v2} 
      \fmf{dbl_wiggly,label=$\mathcal{R}$,tension=2}{H3,t1} 
      \fmf{phantom_arrow,left=0.2}{v1,H1}   
      \fmf{phantom_arrow,left=0.2}{H2,v2}   
      \fmf{fermion,label=$H_u^\dag$,l.side=left}{b1,v1}
      \fmf{fermion,label=$H_d$,l.side=left}{v2,v1}
      \fmf{fermion,label=$H_u$,l.side=left}{v2,b2}
      \fmfdotn{v}{2}
    \end{fmfgraph*}
  }\hfil
  {
    \begin{fmfgraph*}(20,20)   
       \fmfstraight
      \fmfbottom{b1,b2}\fmftop{t1}
      \fmfpolyn{shaded,smooth,pull=1.8,tension=0.5}{H}{3}
      \fmf{phantom}{b1,v1,H1}
      \fmf{phantom}{H2,v1,b2} 
      \fmf{phantom,tension=2.5}{t1,H3}  \fmffreeze
      \fmf{double,label= $\mathcal{X}$,l.side=left,left=0.4}{v1,H1}  
      \fmf{double,label=$\!\!\!\mathcal{X}^\dag$,l.side=left,left=0.4}{H2,v1} 
      \fmf{dbl_wiggly,label=$\mathcal{R}$,tension=2}{H3,t1} 
      \fmf{phantom_arrow,left=0.4}{v1,H1}   
      \fmf{phantom_arrow,left=0.4}{H2,v1}   
      \fmf{fermion,label=$H_u^\dag$ ,l.side=left}{b1,v1}
       \fmf{fermion,label=$\ H_u$,l.side=left}{v1,b2}
       \fmfdotn{v}{1} 
    \end{fmfgraph*}
  }\hfil
  {
    \begin{fmfgraph*}(20,20)   
      \fmfstraight
      \fmfbottom{i1,o1} \fmftop{t1}
      \fmf{fermion,label=$H_u^\dag$,l.side=left}{i1,v1}
      \fmf{fermion,label=$H_u$,l.side=left}{v1,o1}
      \fmf{dbl_wiggly,label=$\mathcal{R}$,l.side=left,tension=2}{t1,v2} 
      \fmf{dbl_wiggly,label=$\mathcal{R}$,l.side=left,tension=2}{v2,v1}      
      \fmfdotn{v}{2}  \fmfblob{.3w}{v2}
    \end{fmfgraph*}
  } 
\end{fmffile}
\caption{Renormalization of the Higgs soft mass operators
  $\mathcal{R}H_u^\dag H_u$. The left-most diagram is due to the
  $\mu$-operator    $\mathcal{X}^\dag H_u H_d$\@.} 
\label{fig:dia2}
\end{figure}
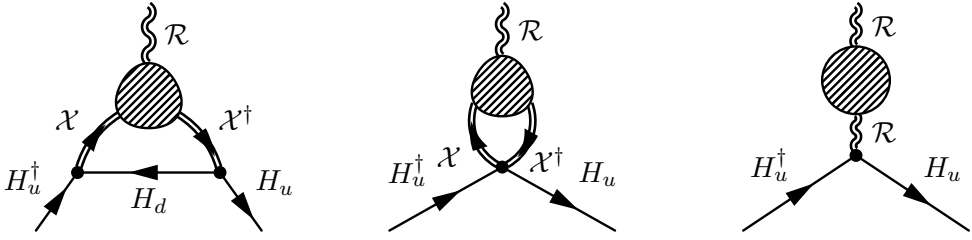
Let us first consider renormalization of the $H_u$ mass. The diagrams
which contribute are the first and third diagrams from left 
in Fig.~\ref{fig:dia2} (according to our definition of the Lagrangian
in Eq.~\eqref{appmuL}). 
Note that neither of the two diagrams involves an internal $H_u$ line.
Thus for the purpose of computing the $H_u$ mass we may treat $H_u$ as a
non-propagating background field. Furthermore, we can drop all
components of $H_u$ except for the scalar. Finally we can even set the
scalar component of $H_u$ equal to a constant. 
Our goal will be to compute the dependence of the low energy
effective Lagrangian on this complex doublet of numbers $H_u$
Our goal will be to compute the dependence of the low energy effective
Lagrangian on $H_u$. The full $H_u$ scalar mass ({\it i.e.} the $(1,1)$
entry of the matrix in Eq.~\eqref{higgs-intro} which consist of both
the SUSY and the non-SUSY contributions) is simply the
coefficient of $ H_u^\dag  H_u$ in this Lagrangian. 
Eq.~\eqref{appmuL} can be rewritten as 
\begin{equation}
\int\!\!\! d^4\theta\ \Bigg[ \left( H_d^\dag + x_\mu \, \Xd H_u \right) 
  \left( H_d + x_\mu^\dag \, \X H_u^\dag \right)  
 + \left( r_u \, \R  - x_\mu^\dag  x_\mu \,
     \frac{\mathcal{X^\dag \! X}}{M_*^2} \right) \!\: 
          H_u^\dag  H_u  \Bigg] \ . 
\end{equation}
Note that there is no kinetic term for the number $H_u$, we have dropped the
operator $\mathcal{R} H_d^\dag H_d$ because it does not contribute to
the renormalization of the $H_u$ mass, and we have grouped the $H_d$
kinetic term and the bosonic part of the $\mu$-operator together by
completing the square.
In order to bring this to a form similar to Eq.~\eqref{appgeneralL},
we redefine the vector of coefficients $r_u$
\begin{equation}
\tilde r_u \equiv r_u -x_\mu^\dag x_\mu \; ,
\label{rshift}
\end{equation}
so that our Lagrangian becomes
\begin{equation}
\int\!\!\! d^4\theta\ \Bigg[ \left( H_d^\dag + x_\mu \, \Xd H_u \right) 
  \left( H_d + x_\mu^\dag \, \X H_u^\dag \right)  +\tilde r_u \, \R
H_u^\dag H_u \Bigg] \; . 
\label{appgeneralL2}
\end{equation}
The $r_u$ redefinition can absorb the couplings of $\mathcal{X^\dag X}$
to $H_u^\dag H_u$ because the vector of real operators $\mathcal{R}$ contains all
possible operators of the form $\mathcal{X^\dag X}$. 
In this basis there is now a one to one correspondence between the terms 
in Eq.~\eqref{appgeneralL2} and the terms in Eq.~\eqref{appgeneralL}. We will
exploit this correspondence when we discuss the diagrammatic proof at the end
of this Section. 

But let us first understand the proof using a field redefinition. We define
\begin{equation}
\tilde H_d \equiv H_d + x_\mu^\dag \, \X \: H_u^\dag \ ,
\label{appredef}
\end{equation} 
and our Lagrangian reduces to
\begin{equation}
\int\!\!\! d^4\theta\ \Bigg[ \tilde H_d^\dag \tilde H_d  +\tilde r_u \, \R
H_u^\dag H_u \Bigg] \; . 
\end{equation}
This field redefinition preserves supersymmetry despite the daggers in
it's definition. The key is that $H_u^\dag$ is not a full
anti-chiral superfield, it is simply a doublet of complex numbers. The
$\mathcal{X}$ appearing in the field redefinition is chiral which is
important because both $\mathcal{X}$ and $H_d$ are dynamical and we do
not want to destroy manifest supersymmetry by mixing up chiral and
anti-chiral fields.

In this new basis things have become very simple. $\tilde H_d$
has completely decoupled and does not contribute to the
renormalization of the $H_u$ mass. The scaling of $\tilde r_u$
entirely comes from the anomalous dimension of $\mathcal{R}$ (the third
diagram in Fig.~\ref{fig:dia2}.)
Thus
\begin{equation}
\tilde r_u \, \Big|_{M_\text{int}} \: = \: 
  \left( \frac{M_\text{int}}{M_*} \right)^\gR 
       \, \tilde r_u \, \Big|_{M_*} \;  .   
\end{equation}
which tends to zero as $M_\text{int} \ll M_*$ because $\gR > 0$. Now
we can read off the $H_u$ scalar mass at $M_\text{int}$. It is
\begin{equation}
\left( \frac{M_\text{int}}{M_*} \right)^\gR  \: 
    \tilde r_u(M_*) \:  \frac{D}{M_*^2}  \quad
      \sim \left( \frac{M_\text{int}}{M_*} \right)^\gR M_\text{SUSY}^2
\end{equation} 
which is negligibly small compared to $M_\text{SUSY}^2$, the mass scale of the
gaugino masses.  Let us emphasize that
this is the full scalar mass which includes both the contribution from
the $\mu$ term as well as from soft supersymmetry breaking.

We can also extract the soft supersymmetry breaking mass $m_{H_u}^2$
by undoing the field redefinition Eq.~\eqref{appredef} at the scale
$M_\text{int}$ to obtain the low-energy Lagrangian (we have set
$\tilde r_u(M_\text{int}) = 0$) 
\begin{equation}
\begin{split}
\int\!\!\!  d^4\theta \: & \left( H_d^\dag + x_\mu \, \Xd
  H_u\right)  \left( H_d + x_\mu^\dag \, \X H_u^\dag \right)  \\ 
 = & \int\!\!\!  d^4\theta \ \Bigg[ H_d^\dag H_d 
    + x_\mu^\dag \, \X  H_d^\dag H_u^\dag 
     + x_\mu \, \Xd H_u H_d  + x_\mu^\dag x_\mu \, 
         \frac{\mathcal{X^\dag \! X}}{M_*^2} H_u^\dag H_u
        \Bigg]  \; .  
\end{split}
\end{equation}
Here the term proportional to $\mathcal{X}$ is the $\mu$-operator
which contributes $+|\mu|^2$ to the $H_u$ scalar mass and the term
proportional to $\mathcal{X^\dag X}$ is the soft mass squared. It is
equal to  
\begin{equation}
m_{H_u}^2 = - \: x_\mu^\dag x_\mu 
    \frac{\mathcal{X^\dag \!  X}}{M_*^2} \Big|_D \: = - |\mu|^2 \; . 
\end{equation}    
\noindent
Of course, with a completely analogous argument we may compute the $H_d$
mass and find 
\begin{equation}
m_{H_d}^2=m_{H_u}^2=-|\mu|^2 \ ,
\end{equation}   
which is what we set out to prove.

Alternatively, we can also construct a diagrammatic proof without
making use of the field redefinition of Eq.~\ref{appredef}. The proof is
completely analogous to the diagrammatic proof considered at the beginning
of the Appendix. Our starting point is the Lagrangian Eq.~\eqref{appgeneralL2},
which leads to the Feynman diagrams in Fig.~\ref{fig:dia2}.
As before, the two left-most diagrams in the Figure have identical blobs and
give canceling contributions to the renormalization of $\tilde r_u$. The
third diagram gives 
\begin{equation}
\frac{d \, \tilde r_u}{dt} = \gR \, \tilde r_u \; .
\end{equation}
For positive $\gR$ this equation has an attractive infrared fixed
point at which $\tilde r_u = 0$. And undoing the shift Eq.~\eqref{rshift}
we obtain $\left| r_u\right|_{M_\text{int}} = \left.  x_\mu^\dag
  x_\mu\right|_{M_\text{int}}$,    
and therefore $m_{H_u}^2=-|\mu|^2$.

\bibliography{reference}
\bibliographystyle{JHEP}

\end{document}